\documentclass[fp,twocolumn]{jpsj3}
\usepackage{graphicx}
\usepackage{dcolumn}
\usepackage{bm}
\usepackage{color}
\begin{document}
%
\title{
Merged Four Dirac Points at the Critical Interlayer Distance \\ 
in Commensurately Twisted
Bilayer Graphene: the Origin of the Zero Velocity
} 
\author{Aya Yamada and Yasumasa Hasegawa}
\inst
{Graduate School of Material Science, 
University of Hyogo, \\
3-2-1 Kouto, Kamigori, Hyogo, 678-1297, Japan \\
\textrm{(Received March 11, 2020; accepted December 3, 2020; published online January 12, 2021)}
}

\abst{
%
 We study the 
commensurately-tilted bilayer graphene in the tight-binding model 
with changing the interlayer distance, which can be tuned by pressure. 
{We find that}
at the commensurately-tilted bilayer graphene with moderate rotating angles, 
when the energy gap at K point is not negligible, 
the other Dirac points within the upper two bands move along the 
$\Gamma$-$K$-$M$ line in the bilayer Brillouzone,
when the interlayer distance is changed. 
The velocity at K point becomes zero due to the merging 
of the four Dirac points
within upper two bands. 
This mechanism of zero velocity at K point
is expected to be the origin of
  the magic angle 
  with flat band at ambient pressure, 
at which 
the upper two bands are almost degenerate and 
the band gap at K can be neglected. 
%
}

\maketitle

\section{Introduction}
Single layer graphene has a two-dimensional honeycomb structure with two sites in the unit cell.
Two bands touch at the corners of the first Brillouin zone (K and K$'$ points) 
in single layer graphene.
{\cite{Neto2009}}
Although a small band gap may open due to the spin-orbit coupling, 
it is very small and can be neglected.
{Spacial modulations of the potential are 
discussed to be important for Dirac fermions.\cite{Gibertini2009, Downing2017}}
Recently, bilayer graphene with a small rotating angle (twisted bilayer graphene)
attracts a lot of interest.
\cite{Rozhkov2016} 
The twisted bilayer graphene has a large unit cell 
as the twist angle becomes 
small.
The velocity at K point is predicted to be zero at the magic angles\cite{Lopes2007,%
Bistritzer2011,Morell2010,Po2018,Hejazi2019}.
It is also shown that the bands at the charge neutral point are exactly flat
when the interlayer coupling is finite only between the sites belonging the different sublattice 
in each layers\cite{Tarnopolsky2019}. 

{When the bands become so flat that the band width or kinetic energy
is in the same order as or smaller than the interaction energy between electrons,
the interaction between electrons becomes important.}
Indeed,  strongly correlated insulator phase 
and superconductivity in twisted bilayer graphene 
at the first magic angle ($\sim 1^{\circ}$)  have been observed\cite{Cao2018,Cao2018b}. 
The magic angle is predicted to be controlled by pressure\cite{Carr2018},
and shown experimentally\cite{Yankowitz2019}.
The zero velocity at K point is a necessary condition for the flat band
but not a sufficient condition.
The study of the mechanism of zero velocity, however,
is important to underestand the magic angle in bilayer graphene. 
In this paper we study the condition for the zero velocity at K point.

%
 \begin{figure}[tb]
\vspace{0.3cm}
\centering 
\includegraphics[width=0.45\textwidth]{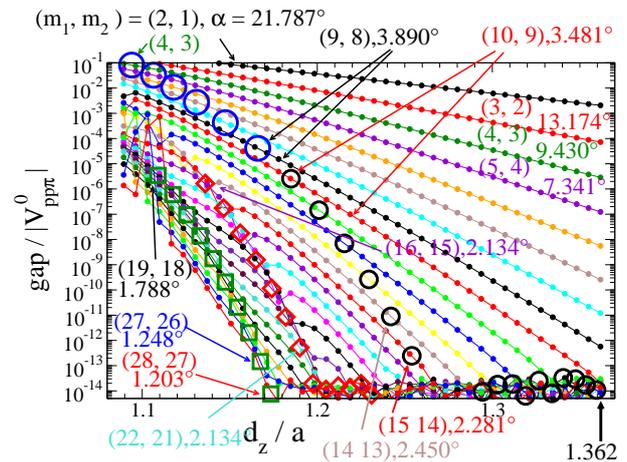}  
\caption{(color online)
Energy gap as a function of the interlayer distance for commensurately rotated angles
with $(m_1,m_2) = (2,1), (3,2), (4,3), \cdots, (28,27)$. The gaps smaller than $\sim 10^{-14}$
are caused by numerical errors and have no serious meanings. 
Large open black circles, red diamonds, and green squares are 
the first, second and third critical interlayer distances, 
where 
the velocity at K with $\delta_k=0.01$ in  Eq.~(\ref{eqdefvelocity}) 
is zero as  V-shaped cusps
which are related to the 
first, second and third magic angles for the commensurately rotated bilayer graphene, 
respectively. 
Large blue circles are the critical interlayer distance, where 
the velocity at K with $\delta_k=0.01$ in  Eq.~(\ref{eqdefvelocity}) 
is zero as a reversed V-shaped cusp.
At the ambient pressure $d_z/a=1.362$.
}
\label{figfiggap} 
\end{figure}
 Many theoretical studies have been done,
\cite{Downing2017,Rozhkov2016,Lopes2007,%
Bistritzer2011,Morell2010,Po2018,Hejazi2019,Tarnopolsky2019,%
Carr2018,%
Shallcross2008,
Mele2010,Shallcross2010,Uchida2014,Sboychakov2015, Rozhkov2017,
Wolf2019}
but there exist a lot of mysteries remain
 to be revealed,
 especially, the mechanism of zero velocity at the magic angles. 
Most of the previous studies have been done in the continuous 
model.
{\cite{Bistritzer2011}}
In the continuous model the unit cell is infinite. 
The energy gap at K point in the moir\'{e} Brillouin zone is zero.
The continuous model can be justified only in the case of small twist angles.
On the other hand,
 the commensurately tilted bilayer graphene has a finite number of
sites in the unit cell (see Appendix). 
The twisted bilayer graphene in the ambient pressure has the fixed interlayer distance,
and the strength of the interlayer coupling with respect to the hopping integrals in the layer
is constant. 
Then the band gap at the Dirac points is 
negligibly small in small twisted angles.
%
\begin{figure}[bt]
\begin{center}
\vspace*{1.2cm} 
\includegraphics[width=0.4\textwidth]{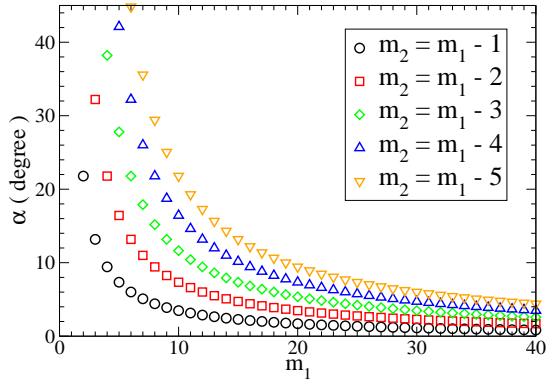} 
\end{center}
\caption{(color online).
Rotating angle $\alpha$ of the commensurately rotated bilayer graphene with $m_1$ and $m_2$
(Eq.~(\ref{eqA13}) and Eq.~(\ref{eqA14})).
When $m_1-m_2$ is divisible by 3 (green diamonds), energy gap is zero\cite{Mele2010}.
In this paper we consider mainly $m_2=m_1-1$ (black circles).  
%
}
\label{figfigcommensurateangle}
\end{figure}
When the twist angle is finite,  a finite gap, however, exists 
  even without the spin-orbit coupling,
as shown in Fig.~\ref{figfiggap}. 
In this figure the rotating angle is taken to be commensurate, i.e., 
the rotating angle $\alpha$ is given by a pair of integers $(m_1, m_2)$, 
as shown in Fig.~\ref{figfigcommensurateangle}.
The energy gap at the Dirac points is caused by 
 the coupling between the distant Dirac points in the Brillouin zone for each layer,\cite{Mele2010}
  which is neglected
in the continuous model. 
Since the vertical axis in Fig.~\ref{figfiggap} corresponds to the strength of the coupling
between K and K$'$,  Fig.~\ref{figfiggap} can be seen as the interlayer-distance and twist-angle dependences
of the coupling strength of the inter-valley scattering.
These dependences should be observed by experiments.

This situation is similar to the zero modes in the presence of the uniform magnetic field
in the single layer graphene with anisotropic hoppings\cite{Hasegawa2006,Esaki2009};
If the hoppings between the nearest sites are anisotropic in the absence
of external magnetic field, the Dirac points moves from
K and K$'$ points, but the energy gap remains zero at Dirac points unless 
two Dirac points merge at one of the time reversal invariant momentums, 
$\Gamma$ and three M points \cite{Hasegawa2012}. 
There exits the zero mode
in a uniform magnetic field in single layer graphene, if the hopping between nearest sites are isotropic. 
The energy gap becomes finite, however, 
if the one of the hoppings is larger than the other two hoppings\cite{Hasegawa2006,Esaki2009}. 
The opening of the gap in the single-layer graphene in magnetic field with the anisotropic
hoppings is shown to be caused by the coupling of two Dirac points\cite{Esaki2009}.

In this paper we study the commensurately twisted bilayer graphene in the tight-binding model 
with changing the interlayer distance. 
We mainly study the case $(m_1, m_2) = (m_1, m_1-1)$.
When $m_1-m_2$ is an integer multiple of 3, 
 the energy gap is zero at $K$ as shown by Mele\cite{Mele2010}. 
 We do not consider that case. 
{We ignore 
the lattice relaxation, which may occur and affect the electron structure 
in the twisted bilayer graphene\cite{Popov2011,Uchida2014,Nam2017}.
}

\begin{figure}[bt]
\begin{center}
\vspace*{1.2cm} 
\includegraphics[width=0.4\textwidth]{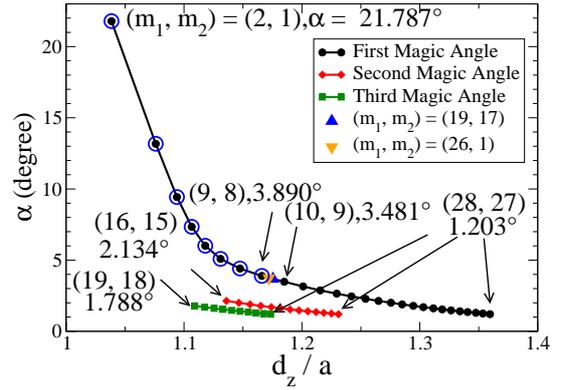} 
\end{center}
\caption{(color online).
First 
(black circles), second 
(red diamonds), and third 
(green squares) 
critical interlayer distances 
in 
 commensurately rotated bilayer graphene  with $m_2=m_1-1$ ($m_1=2, 3, 4, \cdots 28$).
 These points correspond to the first, second, and third magic angles, respectively. 
 Large blue circles show that the
 velocity obtained with $\delta_k=0.01$ in  Eq.~(\ref{eqdefvelocity}) 
is zero as the reversed V-shaped cusp as a function of the interlayer distance. 
The blue triangle and orange downward triangle are 
the critical interlayer distances for 
$m_2=m_1-2=17$ and $m_1=26$ and $m_2=1$ with $\alpha'=60^{\circ}-\alpha$, respectively. 
These two points are on the smooth line of $(m_1, m_1-1)$.
}
\label{figfig0}
\end{figure}

{The energy gap at ambient pressure ($d_z/a =1.362$) is exponentially small
and can be neglected when the rotating angle $\alpha$ is small ($\alpha \lesssim 2.134^{\circ}$).
When the energy gap can be neglected, each two bands are almost degenerated 
around K point and four bands touch at K  point,
which is a massless Dirac point 
(number of the bands are doubled, if the spin degree of freedom is taken into account). 
The velocity at K point becomes zero when the interlayer distance is a critical value
(shown by circles in Fig.~\ref{figfiggap}).
We define it as 
the critical interlayer distance, $d_{zc}$, which is related to the magic angle at ambient pressure. 
Even when the energy gap is not negligible, the K point remains a Dirac point
for upper two bands and lower two bands.
In that case  the velocity at K point becomes zero at a critical value of the interlayer distance.  
As we will show in section \ref{section3}, 
we numerically obtain the velocity by taking the differentiation with 
small $\delta_k$ which we take to be $0.01$. 
When we take a fixed value of $\delta_k$, the  interlayer-distance dependence of the 
velocity at K point changes from a V-shaped cusp 
to a reversed V-shaped cusp.  
}
Even with the finite energy gap at K point,  the velocity at K point
in commensurately twisted bilayer graphene becomes zero 
at the critical value of the interlayer distance,
which we show depend continuously on the interlayer distance, as shown in Fig.~\ref{figfig0}.
We find that 
the zero-velocity 
is caused by the merging of three moving Dirac points and
the fixed Dirac point at K.
If we take smaller value of $\delta_k=0.001$ for example, 
the critical value of the interlayer distance
at $(m_1,m_2)=(10,9)$,
which is identified 
as the V-shaped cusp
with $\delta_k=0.01$,
is shown 
to be the reversed V-shaped cusp.
We conclude that 
when the gap is not negligible, 
the cusp at the critical interlayer distance
is actually 
a reversed V-shaped cusp 
caused by the
merging of four Dirac point at K point. 
%
%
\section{Tight-binding Model in Commensurately Twisted Bilayer Graphene}
We study the tight-binding model in the commensurately twisted bilayer graphene.
{The contents of this section have been discussed in existing literature\cite{Moon2012,Hasegawa2013}.   
}     
The details are given in Appendix \ref{appendixA}.
The rotating angle $\alpha$ is given by two integers $m_1$ and $m_2$ as
\begin{equation}
 \alpha = \arccos \left( \frac{m_1^2+4 m_1 m_2 + m_2^2}{2(m_1^2+m_1m_2+m_2^2)} \right).
\end{equation}
The number of sites in the unit cell is
\begin{equation}
 4 n_0 = 4 (m_1^2 + m_1 m_2 + m_2^2),
\end{equation}
i.e., $n_0$ A sites and $n_0$ B sites  in the first  layer, 
and  $n_0$ A sites and $n_0$ B sites  in the second  layer. 
We plot $\alpha$ as a function of $m_1$ in Fig.~\ref{figfigcommensurateangle}.

The Hamiltonian is given by 
\begin{equation}
 \mathcal{H} = - \sum_{i,j} t(\mathbf{r}_i, \mathbf{r}_j) c_i^{\dagger} c_j,
\end{equation}
where $\mathbf{r}_i$ and $\mathbf{r}_j$ are the lattice sites 
in the commensurately twisted bilayer graphene.
The hopping matrix elements are given by
\begin{align}
 t(\mathbf{r}_i, \mathbf{r}_j)   
=& 
 V_{pp\pi}^0 \exp \left( - \frac{d -a_0}{\delta} \right) \left(1-\left(\frac{d_z}{d} \right)^2\right) 
 \nonumber \\
  &+  V_{pp\sigma}^0 \exp \left(-\frac{d-d_0}{\delta}\right) \left(\frac{d_z}{d}\right)^2 ,
\end{align}
where
$ V_{pp\pi}^0  \approx -0.27$eV is the transfer integral between the nearest sites in the
same layer (the distance between the nearest atoms is $a_0=a/\sqrt{3} \approx 0.142$nm),  
$ V_{pp\sigma}^0  \approx 0.48$eV is the transfer integral between the
atoms in different layers with the same $x, y$ coordinates at ambient pressure
(the distance between the atoms is $d_0 \approx 0.335$nm $\approx 1.362 a$),
the decay length of the transfer integrals is $\delta \approx 0.184 a$,
 and $d = \lvert \mathbf{r}_i - \mathbf{r}_j \rvert$ is the distance between $\mathbf{r}_i$ and $\mathbf{r}_j$. 
We take the interlayer distance $d_z$ as a variable parameter, which can be changed
by pressure. 
Note that $d=\sqrt{d_z^2+d_{xy}^2}$, if  $\mathbf{r}_i$ and $\mathbf{r}_j$ are in the different layers,
where $d_{xy}$ is the distance projected on the plane.
Interlayer transfers are large between atoms with $d_{xy} \ll d_z$, 
when the interlayer transfers are approximately given by
\begin{equation}
 t(\mathbf{r}_i, \mathbf{r}_j) \approx V_{pp\sigma}^0 \exp \left( - \frac{d_z-d_0}{\delta} \right).
\end{equation}
The energy gap at K point is nearly 
proportional to $\exp (-d_z)$ for small $d_z$ as seen in Fig.~\ref{figfiggap}.

In this paper we take $a$ as a unit of the length, 
{$\lvert V_{pp\pi}^0\rvert$} 
as a unit of energy, and
{$\lvert V_{pp\pi}^0\rvert a/\hbar$}
as a unit of velocity.
Therefore, the velocity at the K point in a single layer graphene with only nearest-neighbor hopping,
{$\frac{\sqrt{3}}{2} \lvert V_{pp\pi}^0\rvert a/\hbar$},
is $\frac{\sqrt{3}}{2}= 0.866$
in the unit of 
{$\lvert V_{pp\pi}^0\rvert a/\hbar$}.

\section{Energy Gap and the Velocity at K Point}
\label{section3}
Energy gap at K point in the commensurately rotated bilayer graphene is 
obtained numerically as $\epsilon_{2n_0+1}(\mathbf{K}) - \epsilon_{2n_0}(\mathbf{K})$,
where $\epsilon_{2n_0+1}(\mathbf{K})$ and $\epsilon_{2n_0}(\mathbf{K})$
are the energy at K point in the $(2n_0+1)$th band and the $2n_0$th band
from the bottom, respectively. We plot them as a function of the interlayer distance $d_z$ in 
Fig.~\ref{figfiggap}.
When the twist angle is small ($\alpha \lesssim 3.481^{\circ}$,
i.e. $(m_1, m_1-1)$ with $m_1 \geq 10$)
and $d_z/a$ is near the value at ambient pressure ($d_z/a=1.362$),
 the energy gap at K point is negligibly small 
(smaller than the numerical error $\sim 10^{-14}$) as shown in Fig.~\ref{figfiggap}.
 
 %
\begin{figure}[bt]
\flushleft{(a)  \\} \vspace{0.8cm}
\centering 
\includegraphics[width=0.4\textwidth]{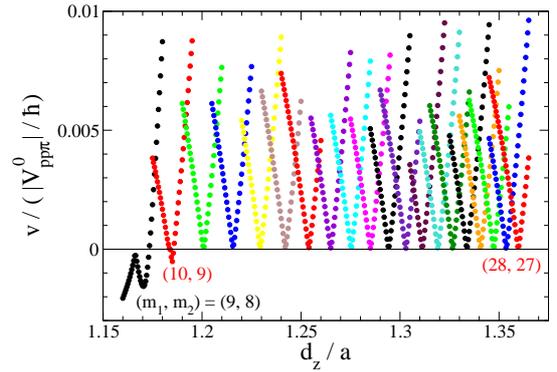}  \\ 
\hfill\flushleft{(b)  \\} \vspace{0.8cm}   
\centering 
\includegraphics[width=0.4\textwidth]{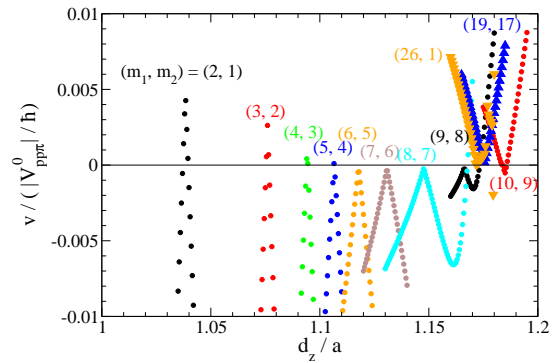}  
\caption{(color online)
Velocities in the $(2n_0+1)$th band at K point defined in Eq.~(\ref{eqdefvelocity}) with $\delta_k=0.01$
as a function of the interlayer distance for commensurately rotated angles
 given by $(m_1, m_2) = (9,8), (10,9), \cdots, (28,27)$  (a) and 
$(m_1, m_2) = (2,1), (3,2), \cdots, (9,8)$ (b). 
Velocities at K point for the commensurately rotated angles given by 
$(m_1, m_2) = (19, 17)$ (blue triangles) 
and $(26, 1)$ orange downward triangles) are also plotted in (b).
}
\label{figfigvelocity} 
\end{figure}
%
%
We calculate the velocity of the $(2n_0+1)$th band at K point as 
\begin{equation}
  v = \frac{\epsilon_{2n_0+1} ((1+\delta_k )\mathbf{K} ) - 
\epsilon_{2n_0+1} (\mathbf{K}) } 
{\delta_k  \lvert \mathbf{K} \rvert}, 
\label{eqdefvelocity}
\end{equation}
where  $\epsilon_{2n_0+1} ((1+\delta_k )\mathbf{K})$
is the energy of the $(2n_0+1)$th band (the band just above the half) at 
the wave number $(1+\delta_k)\mathbf{K}$,
$\mathbf{K}$ is the wave vector of K point in commensurately-twisted bilayer graphene,
{and $\delta_k$ is a dimensionless parameter for the numerical differentiation.
We take $\delta_k=0.01$ in the most part of the paper and take a small value in some cases.}
Since $\frac{3}{2} \mathbf{K}$ is  one of the M points in the extended zone scheme and
$\lvert \mathbf{K} \rvert \equiv \lvert \mathbf{K}-\mathbf{\Gamma} \rvert 
= 2 \lvert \mathbf{M}-\mathbf{K} \rvert$, $v$ is also given by
\begin{equation}
v= \frac{\epsilon_{2n_0+1} (\mathbf{K} + 2 
\delta_{k} 
(\mathbf{M}-\mathbf{K})) - 
\epsilon_{2n_0+1} (\mathbf{K}) } 
{2 \delta_{k}  
\lvert \mathbf{M} - \mathbf{K} \rvert} .
\label{eqdefvelocity2}
\end{equation}
In fig.~\ref{figfigvelocity} we plot the velocities defined by Eq.~(\ref{eqdefvelocity}) with $\delta_k=0.01$ as
functions of $d_z/a$ for $(m_1, m_1-1)$ ($m_1=2,3, \cdots, 28$), $(m_1,m_2)=(19,17)$,
and $(m_1,m_2)=(26,1)$. 
In our choice of parameters, the velocity at K point (the Dirac point) in the bilayer 
graphene is zero when $m_1=28$ and $m_2=27$  at ambient pressure 
when $d_z/a=1.362$, where $a$ is the lattice constant in each layer 
and $d_z$ is the distance between layers. 
The result at $m_1=28$ and $m_2=27$  is consistent with
the previous results in the tight-binding approximation\cite{Morell2010}
and the continuous approximation\cite{Lopes2007,%
Bistritzer2011,Morell2010,Hejazi2019,Tarnopolsky2019}. 
We also find that the second and the third magic angles are obtained in the tight-binding 
model when the interlayer distance becomes small.

As seen in Fig.~\ref{figfigvelocity}, there is a qualitative difference between
the $d_z$-dependences of $v$ near $v=0$ between those with $m_1 \geq 10$ ($m_2=m_1-1$) and
those with $m_1 \leq 9$ ($m_2=m_1-1$).   
The velocity  of the upper ($(2n_0+1)$th) band at K point is positive near the critical value
of $d_z$ when $m_1 \geq 10$, (V-shaped cusp), while that is negative when    $m_1 \leq 9$
(reversed V-shaped cusp).
%
\begin{figure}[bt]
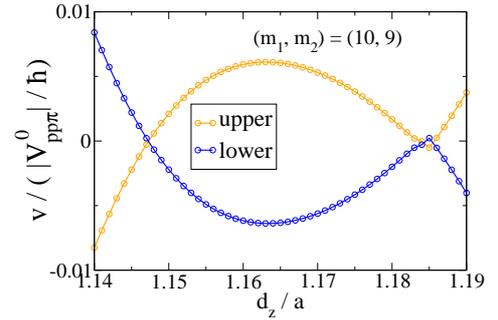
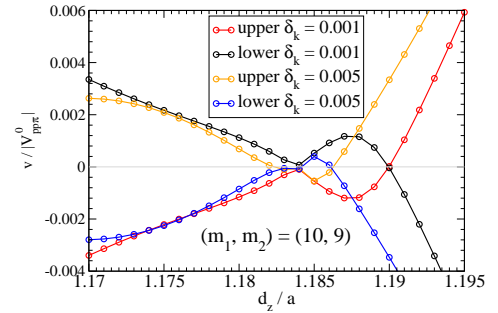

\centering 
%
\flushleft{(a)\\} \vspace{0.7cm}
\centering
\includegraphics[width=0.35\textwidth]{69657fig5a.eps}   
\flushleft{(b)\\} \vspace{0.7cm}
\centering
\includegraphics[width=0.35\textwidth]{69657fig5b.eps}   
%
\flushleft{(c)  \\} \vspace{0.7cm}
\centering
\includegraphics[width=0.35\textwidth]{69657fig5c.eps}   
\caption{(color online)
Velocity in the upper ($(2n_0+1)$th) band (orange) and the lower ($2n_0$th) band (blue) at K point 
as a function of the interlayer distance for commensurately rotated angles
 given by$(m_1, m_2) = (9,8)$ [(a)] and $(10,9)$ [(b) and (c)]. We take $\delta_k=0.01$ in (a) and (b).
 In (c) $\delta_k=0.001$ and $0.005$ are taken. The V-shaped cusp remains for $\delta_k=0.005$
 (orange circles and line), but the reversed V-shaped cusp appears for $\delta_k=0.0001$
 (red circles and line), which is similar to (a) with $(m_1,m_2)=(9,8)$.
 As $\delta_k$ becomes smaller, the value of  sign-change  at $d_z/a \approx 1.147$ for $\delta_k=0.01$
 [(b)] comes to 
{$d_z/a \approx 1.184$} 
and disappears for $\delta_k=0.001$.
 The V-shaped cusp with a negative minimum value of $v$ at $d_z/a \approx 1.185$ 
 for $\delta_k =0.01$
 [orange points and line in (b)] becomes a reversed V-shaped cusp at $d_z/a \approx 1.184$
with the maximum  value of $v=0$ for $\delta_k=0.001$ [red circles and line in (c)].
}
\label{figvvsdz98109} 
\end{figure} 
We plot the velocity of the $(2n_0+1)$ band (upper band) calculated by Eq.~(\ref{eqdefvelocity}) and
the velocity of the $2n_0$th band (lower band) calculated by
\begin{equation}
  v' = \frac{\epsilon_{2n_0} ((1+\delta_{k} )\mathbf{K} ) - 
\epsilon_{2n_0} (\mathbf{K}) } 
{\delta_{k}  \lvert \mathbf{K} \rvert}, 
\label{eqdefvelocitylow}
\end{equation}
for $(m_1, m_2) = (9,8)$ and $(10,9)$ in Fig.~\ref{figvvsdz98109}.
The velocities of the upper band $v$ and the lower band satisfies $v=-v'$ in these regions.
We discuss the $(m_1, m_2) = (9,8)$ case and $(10,9)$ case separately in 
following subsections. The case of $(m_1, m_2) = (2,1)$ is also discussed.

\subsection{$(m_1, m_2) = (9,8)$}
\begin{figure}[bt]
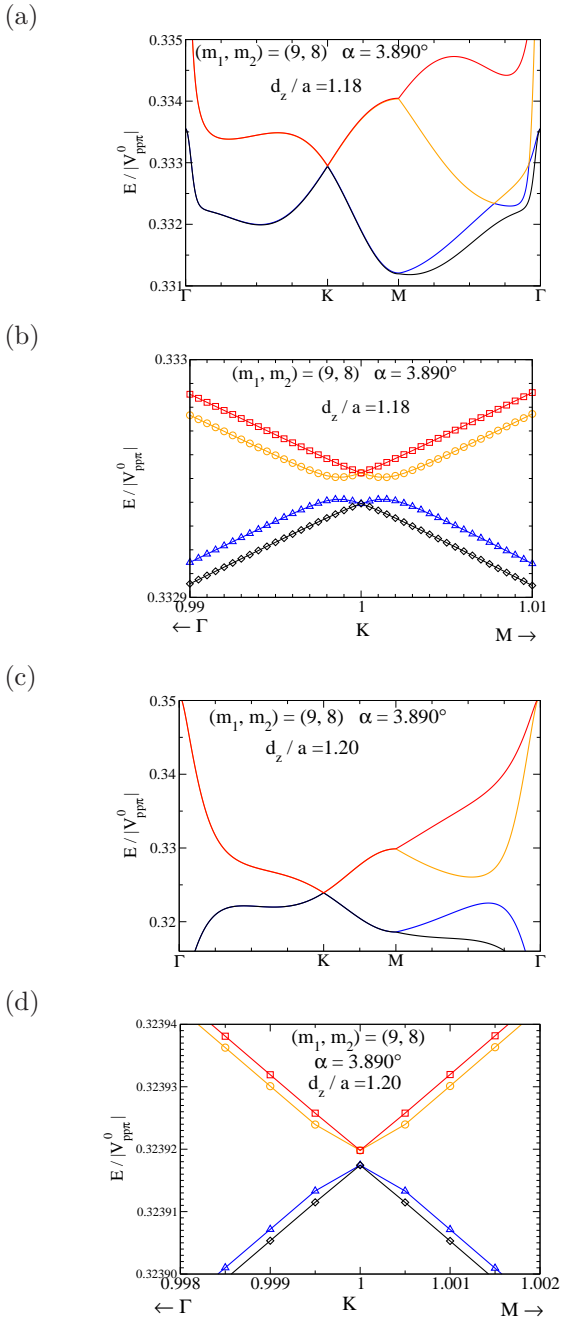

\centering 
\flushleft{(a)  \\} \vspace{0.cm}
\centering
\includegraphics[width=0.32\textwidth]{69657fig6a.eps}   
\flushleft{(b)  \\} \vspace{0.cm}
\centering
\includegraphics[width=0.32\textwidth]{69657fig6b.eps} \\  
%
\flushleft{(c)  \\} \vspace{0.cm}
\centering
\includegraphics[width=0.32\textwidth]{69657fig6c.eps}   
\flushleft{(d)  \\} \vspace{0.cm}
\centering
\includegraphics[width=0.34\textwidth]{69657fig6d.eps}   
\caption{(color online)
Energy dispersion in the commensurately twisted bilayer graphene with $(m_1, m_2)=(9,8)$.
The interlayer distance is $d_z/a=1.18$ [(a) and (b)], and $d_z/a=1.20$ [(c) and (d)].
At K point the finite gap is seen in the close-up figures (b) and (d),
where the momentum is scaled by $\lvert \mathbf{K} \rvert$. 
The degeneracy of the bands is lifted. The Dirac point at K is separated to two Dirac points
{of upper two bands ($(2n_0+1)$th and $(2n_0+2)$th bands) and
lower two bands ($(2n_0-1)$th and $2n_0$th bands)}. 
}
\label{figband98118} 
\end{figure}

As seen in Fig.~\ref{figvvsdz98109}~(a) (orange curve), $v<0$ at $d_z/a \lesssim 1.173$.
This value of the sign-change point,
 however, is not a intrinsic one, but depends on the choice of $\delta_k$.
 As shown in Fig.~\ref{figband98118}~
(b) and (d), 
the Dirac point at K is not the degenerated point of four bands. The Dirac 
point is separated into two Dirac points of upper two bands and lower two bands
due to the finite gap at K point,  
which has been shown by Mele\cite{Mele2010}. 
As a result the velocity of the upper band at K point,
whichi is positive when we take $\delta_{k}=0.01$
 (orange curve in Fig. \ref{figvvsdz98109}(a)), 
is negative due to the small but finite energy gap when $d_z/a=1.18$,
if we take smaller value of $\delta_k$. Therefore,
the sign-change point at  $d_z/a \sim 1.173$ becomes larger, if we 
calculate the velocity using a smaller value of $\delta_k$. 
 Even when $d_z/a=1.20$, there exists a finite gap at K point,
as seen in Fig.~\ref{figband98118}~(d).
Therefore, the sign-change point at $d_z/a \sim 1.173$ is not intrinsic.
In principle, the velocity of K point is negative (or zero) in the limit of $\delta \to 0$,
if a gap between $(2n_0)$th band and $(2n_0+1)$th band cannot be neglected
and there exist massless Dirac points between $(2n_0+1)$th band and $(2n_0+2)$th band.
%

\begin{figure}[bt]
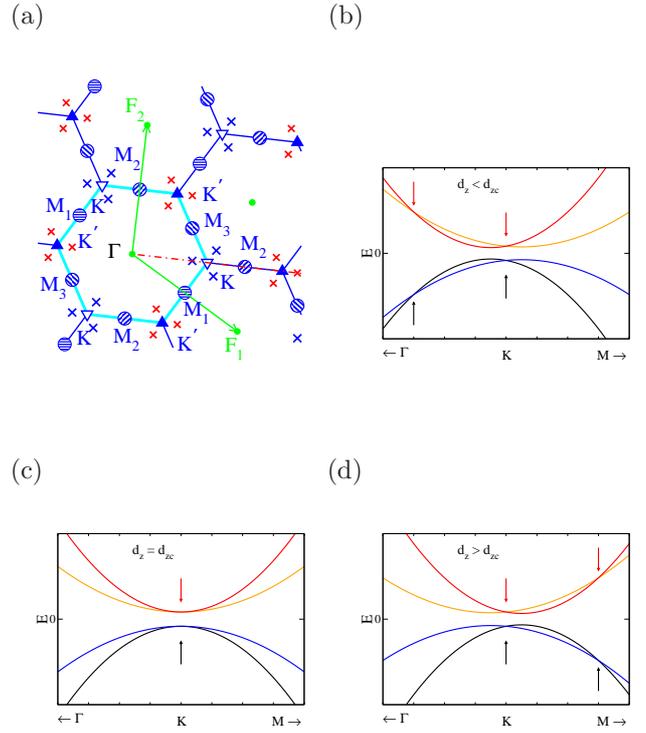

\flushleft{(a)  \hspace{3.5cm} (b) \\  } 
\vspace{6mm}
\centering
\includegraphics[width=0.2\textwidth]{69657fig7a.eps} 
\hspace{5mm} 
\includegraphics[width=0.2\textwidth]{69657fig7b.eps} 
              \vspace{1cm} \\ 
\flushleft{(c) \hspace{3.5cm} (d) \\} %
\vspace{6mm}
\centering 
\includegraphics[width=0.2\textwidth]{69657fig7c.eps} 
\hspace{5mm}
\includegraphics[width=0.2\textwidth]{69657fig7d.eps} \vspace{1cm} 
\caption{(color online)
(a): When the interlayer distance is close to the critical value, $d_{zc}$,
there are six Dirac points (three red crosses and three blue crosses)
 in the first Brillouin zone (cyan hexagon) of the bilayer graphene. 
Dirac points at K (blue triangular down) and $K'$ (filled blue triangular up)
 have topological number $+1$ and $-1$, respectively. 
(b), (c), and (d): Schematic figures of the energy band near $K$ point
are shown for $d_z < d_{zc}$, $d_z=d_{zc}$, and $d_z > d_{zc}$, respectively.
At $d_z = d_{zc}$ four Dirac points merge at $K$.
}
\label{schematicfig}
\end{figure}

\begin{figure}
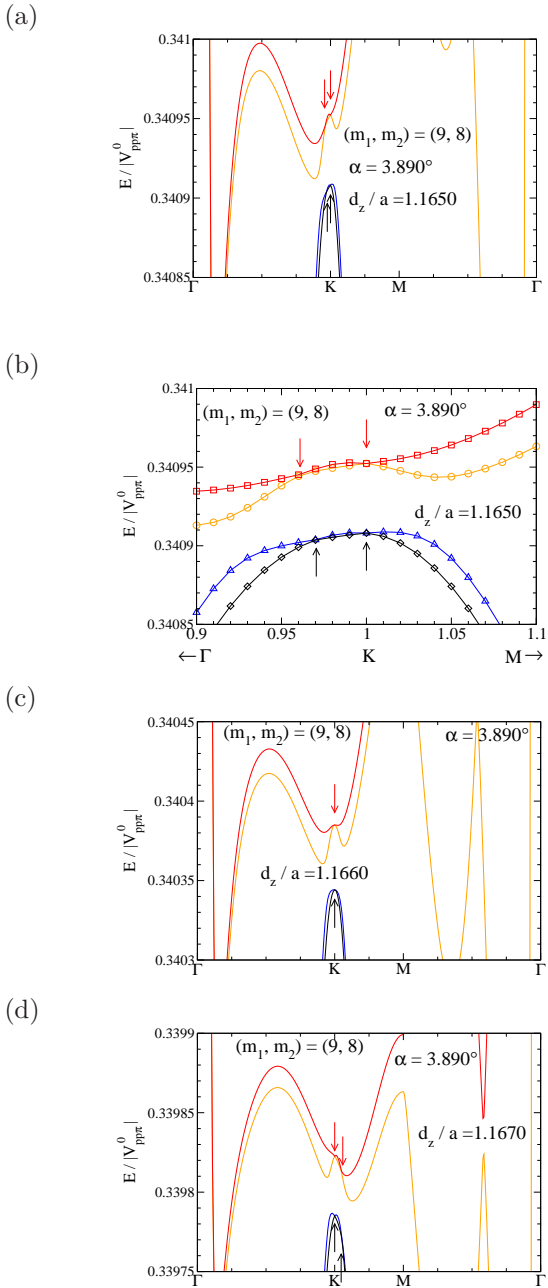

\flushleft{(a)  \\} 
\centering 
\includegraphics[width=0.32\textwidth]{69657fig8a.eps}  
\vspace{5mm}
\flushleft{(b)  \\} 
\centering 
\includegraphics[width=0.32\textwidth]{69657fig8b.eps} \\ 
\flushleft{(c) \\} 
\centering 
\includegraphics[width=0.32\textwidth]{69657fig8c.eps}  
\flushleft{(d) \\ } 
\centering 
\includegraphics[width=0.32\textwidth]{69657fig8d.eps}  \\
\caption{(color online)
Energy dispersion in the commensurately rotated bilayer graphene
with $(m_1, m_2) = (9,8)$ 
near the critical value of $d_{zc}=1.1660 a$. 
 The black, blue, orange, and red lines are the
$(2n_0-1)$th, $(2n_0)$th, $(2n_0+1)$th, and $(2n_0+2)$th bands from the bottom, respectively.
The interlayer distances are taken as  $1.165 a$ [(a) and (b)], $1.166 a$ [(c)], and $1.167 a$ [(d)].
Red arrows indicate the Dirac point in the $(2n_0+1)$th and $(2n_0+2)$th
bands, and black arrow indicate  the Dirac point in the $(2n_0-1)$th and $(2n_0)$th
bands.
}  
\label{figfigband98} 
\end{figure}

\begin{figure}
\vspace{0.5cm}
\centering 
\includegraphics[width=0.45\textwidth]{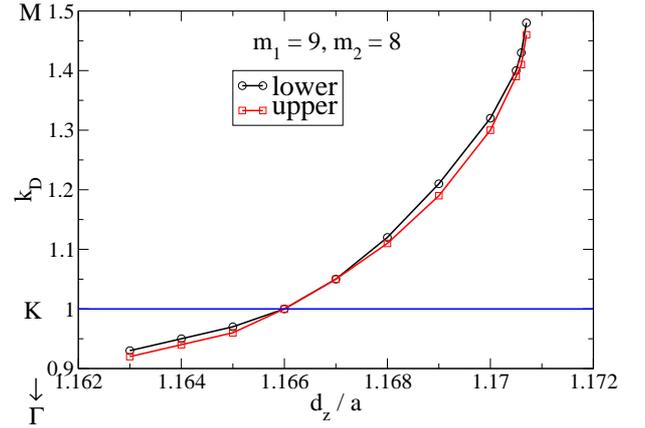}  
\caption{(color online)
The moving Dirac points 
for the commensurately rotated bilayer graphene with $(m_1, m_2)= (9,8)$ 
as a function of interlayer distance.
Four Dirac points merge at K point at $d_z/a \sim 1.1660$.
Two Dirac points merge and annihilate at M point at $d_z/a \sim 1.1707$. 
}
\label{figfigkvsdz} 
\end{figure}

On the other hand, the velocity of the upper band depends on the interlayer  
distance as the reversed V-shaped cusp around $d_z/a \sim 1.166$ 
in the case of  $(m_1, m_2)=(9,8)$ 
(orange curve in Fig.~\ref{figvvsdz98109}~(a)), which is also seen in the case of 
 $m_1 \leq 9$, $m_2=m_1-1$ 
(see Fig.~\ref{figfigvelocity}). 
The reversed V-shaped cusp 
can be understood as follows. 
When the interlayer distance is close to the 
critaical value ($d_{zc}/a \approx 1.166$ in the case of $(m_1,m_2)=(9,8)$)
 there exit other Dirac points
on the lines of $\Gamma$-K-M in the extended zone scheme of Brillouin zone
(See Fig.~\ref{schematicfig}).
When $d_z/a \lesssim 1.166$  or $d_z/a \gtrsim 1.166$,
three Dirac points exist near K point, as shown in Fig.~\ref{figfigband98} (a), (b)
and (d).
The Dirac points move as shown in  Fig.~\ref{figfigkvsdz}. 
At $d_z/a =d_{zc}/a \approx 1.166$
four Dirac points (three moving Dirac points and one fixed Dirac point) meet at K point
(Fig.~\ref{figfigband98} (c)),
and at $d_z/a \approx 1.1704$ two moving Dirac point merge at M point and they disappear at
$d_z/a \gtrsim 1.1704$.  

Due to the $2\pi/3$ rotational symmetry and the time reversal symmetry
in the twisted bilayer graphene, besides the Dirac points at K and K$'$ points, 
six Dirac points should exist on  three  $\Gamma-K$ and  three $\Gamma-K'$ lines in the 
first Brillouin zone in the twisted bilayer graphene, if at least one Dirac point exists
in one of the the $\Gamma-K$ or $\Gamma-K'$ lines 
(see Fig.~\ref{schematicfig}). 
As the interlayer distance 
are changed to the critical value gradually, 
the Dirac points move to K point on the $\Gamma-K$ line 
and finally three moving Dirac points meet at the
$K$ point.
The topological number (Berry phase) of the fixed Dirac point at K is $+1$ and that at K$'$
is $-1$ and that of moving Dirac points to K (K$'$) are $-1$ ($+1$). 
Therefore, at the critical value of the interlayer distance,
where four Dirac points meet at K and K$'$, the topological number is $-2$ and $+2$ at K and K$'$, respectively. 
As a result annihilation of the Dirac points cannot happen in this case.
Further change of the interlayer distance makes one Dirac points at K point and three 
moving Dirac points on the K-M$_j$ line, where $j=1, 2$ and $3$. 
Note that $\mathrm{K}-\mathrm{M}_j$ line lies on the same line in $\Gamma-\mathrm{K}$ line in the 
 extended zone scheme.
This situation is the same as the merging of Dirac points in the single-layer
graphene with isotropic nearest-site hopping $t$ and isotropic third-nearest-site hopping
$t_3 \approx \frac{1}{2} t$\cite{Hasegawa2012}.
Since the topological number is $\pm 2$ at the critical value of the interlayer distance,
the energy depends as
\begin{equation}
 \epsilon (\mathbf{K} + \mathbf{k}) \propto \lvert \mathbf{k} \rvert^2,
\end{equation}
and the velocity at K point becomes zero. 
Therefore, we obtain that $v$ shows the reversed V-shaped cusp with the maximum value 0
as a function of $d_z/a$.

\subsection{$(m_1,m_2)= (10,9)$}
The positions of sign-change and the cusp of velocity are exchanged in 
$(m_1, m_2)=(9,8)$ (Fig.~\ref{figvvsdz98109}~(a)) and $(m_1, m_2)=(10,9)$ (Fig.~\ref{figvvsdz98109}~(b)). 
The velocity of the $(2n_0+1)$th band at K point is negative in a
narrow region near $d_z/a \approx 1.185$ with 
V-shaped cusp when $(m_1, m_2)=(10,9)$,
as seen in Fig.~\ref{figfigvelocity}  and Fig~\ref{figvvsdz98109}(b) orange curve,
where we take $\delta_k=0.01$. The velocity changes sign and become negative
at $d_z/a \lesssim 1.147$.
The value of the sign-change point $d_z/a \approx 1.147$ is not intrinsic, as in the case 
of $(m_1, m_2)=(9,8)$. 
The V-shaped cusp also depends on $\delta_k$, as shown in  Fig~\ref{figvvsdz98109}(c) 
(orange and red curvs). 
 We 
 study the V-shaped cusp near $d_z/a \approx 1.185$ in detaail.
 %
\begin{figure}
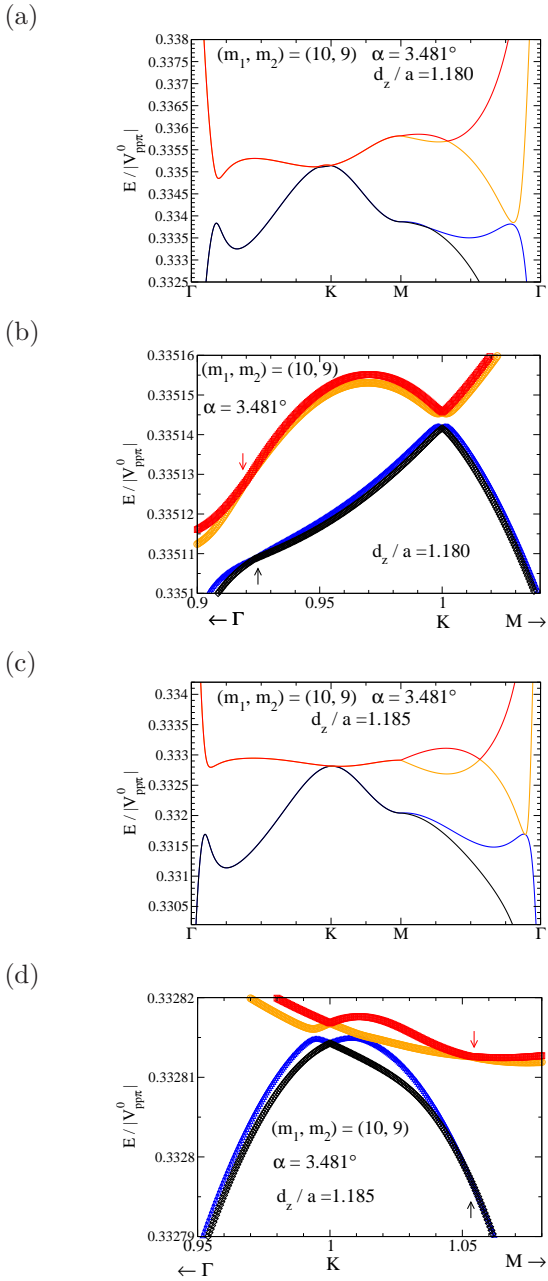

\flushleft{(a)  \\} 
\centering
\includegraphics[width=0.32\textwidth]{69657fig10a.eps}   
\flushleft{(b)  \\} 
\centering
\includegraphics[width=0.32\textwidth]{69657fig10b.eps}  \\
\flushleft{(c)  \\} 
\centering
\includegraphics[width=0.32\textwidth]{69657fig10c.eps}
\flushleft{(d)  \\} 
\centering
\includegraphics[width=0.32\textwidth]{69657fig10d.eps}  
\caption{(color online)
Energy dispersions of $(2n_0-1)$th (black), $2n_0$th (blue), $(2n_0+1)$th (orange),
and $(2n_0+2)$th (red) bands near K point 
at $d_z/d=1.180$ [(a) and (b)] and $1.185$ [(c) and (d)] 
for commensurately rotated angles
 given by
$(m_1, m_2) = (10,9)$. 
Red and blue arrows in (b) and (d) indicate the moving Dirac points.
}
\label{figbandgap0x} 
\end{figure}
We plot the energy dispersion of the bands near half-filling 
at $d_z/a=1.18$ and $1.185$ for $(m_1, m_2)=(10,9)$ in Fig.~\ref{figbandgap0x}.
When we look closer near K point, the small gap between $2n_0$th and $(2n_0+1)$th bands
is seen and the Dirac point separates into two Dirac points on the $\Gamma$-$K$-$M$ line. 
Therefore, if we take $\delta_k$ small enough, the velocity of the $(2n_0+1)$th band at K 
is negative, although it is positive at $d_z/a=1.18$ with $\delta_k=0.01$ 
(see orange curve in Fig.~\ref{figvvsdz98109}~(b)).
There are other Dirac points beside K (near $0.92 \mathbf{K}$ 
in Fig.~\ref{figbandgap0x}~(b), red arrow),
and they move toward M via K as $d_z/a$ increases from $1.180$ to $1.185$. 
This situation is the same as happened near $d_z/a=1.66$ in the case of
 $(m_1, m_2)=(9,8)$ (see Fig.~\ref{figfigband98}).
Therefore, if we take $\delta_k$ small enough,
we expect the reversed V-shaped cusp in the $d_z$-dependence of the
velocity of $(2n_0+1)$th band at K
 in the narrow region
around $d_z/a \approx 1.185$, where the velocity calculated with $\delta_k=0.01$ is negative. 
At the top of the reversed V-shaped cusp, four Dirac points merge and the velocity is zero.

When the rotating angle $\alpha$ is smaller than $3.481^{\circ}$ ($m_1 >10$, $m_2=m_1-1$),
the energy gap at K is small 
and the upper two bands  near K
are almost degenerate each other. 
 As a result,
we have to take $\delta_k$ much smaller to obtain the negative 
region of $v$ and reversed V-shaped
cusp in $d_z$-dependence of $v$. 
If we take $\delta_k$ not small enough,  
we obtain the V-shaped cusps as in Fig.~\ref{figfigvelocity}~(a).

We may expect that the mechanism of
the V-shaped cusp in Fig.~\ref{figfigvelocity}~(a) seen in $(m,m-1)$ with $m \geq 10$ 
is the same as that of the reversed V-shaped cusp with $m \leq 9$;
they are caused by the merging of four 
Dirac points  (three moving Dirac points and the Dirac point at $K$).  
Then, the nearly flat band at the magic angle at ambient pressure
is thought to be the result of the merging of the four Dirac points,
although it is difficult to show explicitly because of the very small energy gap at $K$
and almost degeneracy of $(2n+1)$th and $(2n+2)$th bands. 

\subsection{$(m_1, m_2)=(2,1)$}
In this subsection we discuss the twisted bilayer graphene  with $(m_1, m_2)=(2,1)$
($\alpha=21.787^{\circ}$).
The top of the reversed V-shaped cusp obtained by $\delta_k=0.01$ is positive as seen in 
Fig.~\ref{figfigvelocity}~(b). 
We show that 
the top of the cusp is zero as we take $\delta_k$ smaller.

\begin{figure}
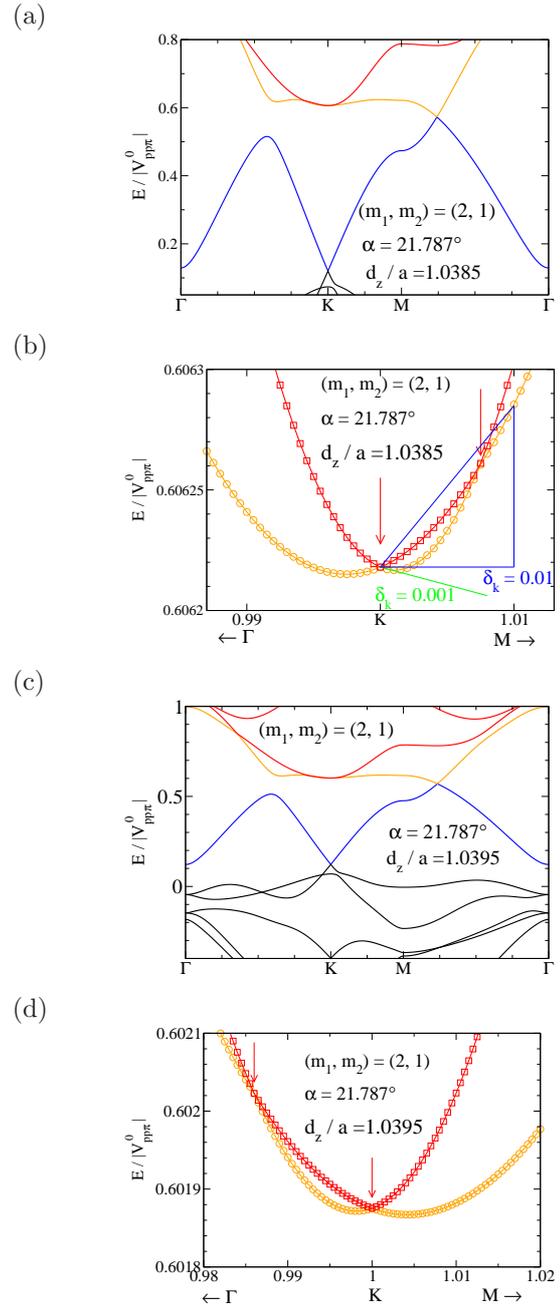

\flushleft{(a)  \\} 
\centering
\includegraphics[width=0.32\textwidth]{69657fig11a.eps} \\
\flushleft{(b)  \\} 
\centering
\includegraphics[width=0.32\textwidth]{69657fig11b.eps}  \\
\flushleft{(c)   \\} 
\centering
\includegraphics[width=0.32\textwidth]{69657fig11c.eps} \\
\flushleft{(d)   \\} 
\centering
\includegraphics[width=0.32\textwidth]{69657fig11d.eps}  
\caption{(color online)
Energy dispersions of $(2n_0-1)$th (black), $2n_0$th (blue), $(2n_0+1)$th (orange),
and $2n_0+2)$th (red) bands 
 for commensurately rotated bilayer graphene with 
$(m_1, m_2) = (2,1)$.
The interlayer distance is $d_z/a=1.385$ [(a) and (b)] and $1.395$ [(c) and (d)].
{(b) and (c) show the energy dispersion near K point. 
The velocity of the $(2n_0+1)$th band at K point is calculated to be  positive
when we take $\delta_{\mathbf{k}}=0.01$ as  blue lines in (b), but it is negative 
when we take $\delta_{\mathbf{k}} \lesssim 0,001$ as shown by green line. }
}
\label{figband2110385} 
\end{figure}
In Fig.~\ref{figband2110385} we plot the energy dispersion for $(m_1,m_2)=(2,1)$ near half filling at the
interlayer distance $d_z/a=1.0385$ and $1.0395$, when the velocity 
of the $(2n_n+1)$th band is small.
Although $v$ at $d_z/a=1.0385$ obtained with $\delta_k=0.01$ is positive, it should be negative 
if we take $\delta_k \lesssim 0.001$, as seen in  Fig.~\ref{figband2110385}~(b).
There exist the moving Dirac points in the K-M line (three moving Dirac points by symmetry),
and four Dirac points meet at K when $d_z/a$ is changed. Therefore, the positive value at the 
top of the reversed
V-shaped cusp around $d_z/a \approx 1.039$ becomes zero when we take smaller $\delta_k$,

If we calculate the velocity at K point in the $(2n_0+2)$th band instead of the $(2n_0+1)$th band,
the interlayer-distance dependence of the velocity in the $(2n_0+2)$th band is always
V-shaped cusp and
the $\delta_{\mathbf{k}}$ dependence is weak. 
The $\delta_{\mathbf{k}}$-sensibility of the velocity makes it possible to distinguish 
the critical values of the interlayer distance 
in the presence of non-negligible band gap and those in the case of negligible band gap.

The moving Dirac point in the case of $(m_1,m_2)=(2,1)$ moves as 
$\mathrm{M} \to \mathrm{K} \to \Gamma$, as $d_z/a$ increases around the reversed V-shaped cusp.
This motion is reversal to those in the cases of $(m_1,m_2)=(9,8)$ and $(10,9)$.
Another curious feature in $(m_1,m_2)=(2,1)$ case is that
the velocity of the $2n_0$th band at K
 does not become small when that of the $(2n_0+1)$th band
is small, as seen in Fig.~\ref{figband2110385}~(a) and (b). 
Since the particle-hole symmetry is broken in the commensurately rotated bilayer graphene,
this asymmetry may happen, but the real reason is not clear.
Despite these curious things, 
the critical interlayer distance 
defined by the zero velocity in the $(2n_0+1)$th band
seems to be a continuous function as a interlayer distance (see Fig.~\ref{figfig0}).

\section{Discussions and Conclusion}
In this paper 
we study the critical interlayer distance in the commensurately twisted
bilayer graphene, 
at which
the velocity of the band just above the half-filling is zero.
We showed that 
the critical interlayer distance 
exists 
 even when the energy gap at K point is not negligible.
 The mechanism of 
the zero velocity 
is shown to be the merging of 
 the four Dirac points at K point. 

In the continuous approximation, the energy gap at K point is neglected.
The energy gap is, however, finite at K point and  the degeneracy of the two bands near K point
is lifted, if we take into account the coupling between K and K$'$ points,
which is in general finite in the tight binding model.
In the tight binding model, the energy gap at K point is very small and can be safely
neglected, only if the twist angle is small,
as shown in Fig.~\ref{figfiggap}. 
We obtain the interlayer-distance dependence 
of 
the velocity at K point 
in the commensurately-twisted bilayer graphene 
with the $4n_0$ cites in the unit cell numerically.

The velocity of the $(2n_0+1)$th band at K as a function of interlayer distance 
becomes zero as a V-shaped cusp when $m_1 \geq 10$ and $m_2=m_1-1$, 
while it becomes zero as a reversed V-shaped cusp  when $m_1 \leq 9$ and $m_2=m_1-1$,
when we calculate the velocity
by the numerical differentiation with $\delta_k=0.01$ instead of taking 
the limit of $\delta_k \to 0$.
By studying the cases with $(m_1, m_2)=(9,8)$ and $(10,9)$ in detail,
we obtain that 
a reversed V-shaped cusp  
is caused by the merging of four Dirac point at K 
in both cases, if we take sufficiently small value of $\delta_k$. 
At 
the critical interlayer distance
one Dirac point with topological number $\pm 1$
 and three moving Dirac points with topological number $\mp 1$ merge at K and K$'$ points,
resulting the topological number $\mp 2$. 
The topological number $\mp 2$ means that two band touch quadratically at K and K$'$ points,
resulting the zero velocity.

This mechanism of the zero velocity at K point is shown to
work 
in the case of small energy gap between $2n_0$the and $(2n_0+1)$th bands with $(m_1,m_2)=(10,9)$,
as well as 
the cases of moderate energy gap 
with $(m_1,m_2)=(9,8)$ and $(2,1)$. 

At small rotating angles, the gap between $2n_0$th and $(2n_0+1)$th bands is
exponentially small, and it can be safely neglected. In that case it is difficult to distinguish
the reversed V-shaped cusp and V-shaped cusp. 
If there is a finite gap,  the velocity at K is negative 
when $\delta_k \to 0$ limit is taken. In that sense 
the zero velocity should be always realized as the reversed V-shaped cusp. 

In this paper we have not concerned the flatness of the band at the magic angle, and
we only study the velocity at K point.
Indeed, the band is not flat 
at the critical interlayer distance, 
when the interlayer distance is small
and 
 the twist angle is not small. 
In the case of a small angle with negligible band gap and almost degenerate upper two bands,
 it is difficult to show explicitly that the flat band is caused by the merging of four Dirac points.
Although we have shown that the zero velocity is caused by the merging of four Dirac points at K
only for the case of  moderate rotation angles, 
this mechanism is thought to  work also for the magic angle with flat band. 



\appendix 
\section{Commensurately Twisted Bilayer Graphene} \label{appendixA}

We take the primitive lattice vectors of the first layer, $\mathbf{a}_1^{(1)}$ and $\mathbf{a}_2^{(1)}$,
as
\begin{align}
 \mathbf{a}_1^{(1)} &= a \begin{pmatrix} \frac{\sqrt{3}}{2} \\ - \frac{1}{2} \end{pmatrix} ,\\
 \mathbf{a}_2^{(1)} &= a \begin{pmatrix} \frac{\sqrt{3}}{2} \\   \frac{1}{2} \end{pmatrix} ,
\end{align}
where $a$ is the lattice constant 
$a=0.246$nm. 
The angle 
between two primitive lattice vectors are $\pi/3$;
\begin{equation}
 \mathbf{a}_2^{(1)} =  R_{\frac{\pi}{3}} \mathbf{a}_1^{(1)} ,
 \end{equation}
 where the 2D rotational matrix $R_{\theta}$ is given by
\begin{equation}
 R_{\theta} = \begin{pmatrix}
 \cos \theta & -\sin \theta \\
 \sin \theta &  \cos \theta
 \end{pmatrix} .
 \end{equation}
%
\begin{figure}[bt]
%
\begin{center}
\includegraphics[width=0.48\textwidth]{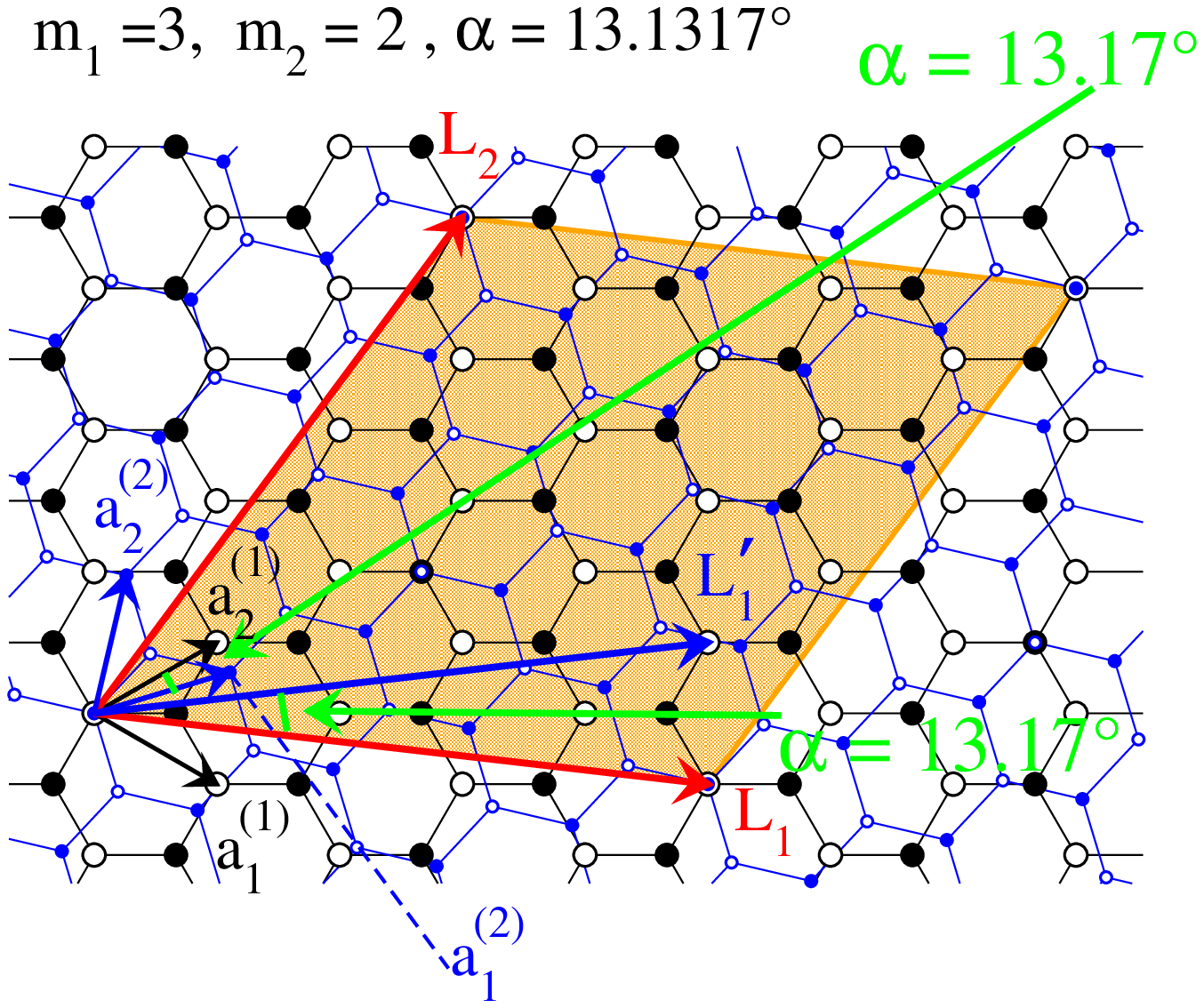} 
\end{center}
\caption{(color online).
Commensurately twisted bilayer graphene with $(m_1,m_2)=(3,2)$.
Large (small) open and filled circles are A and B sublattice in 
the first (second) 
layer, respectively. The second layer is rotated by $\pi/3 -\alpha$.
The orange area is the unit cell of the commensurately twisted bilayer graphene.
}
\label{figfig32}
\end{figure}
\begin{figure}[tb]
\centering 
\includegraphics[width=0.47\textwidth]{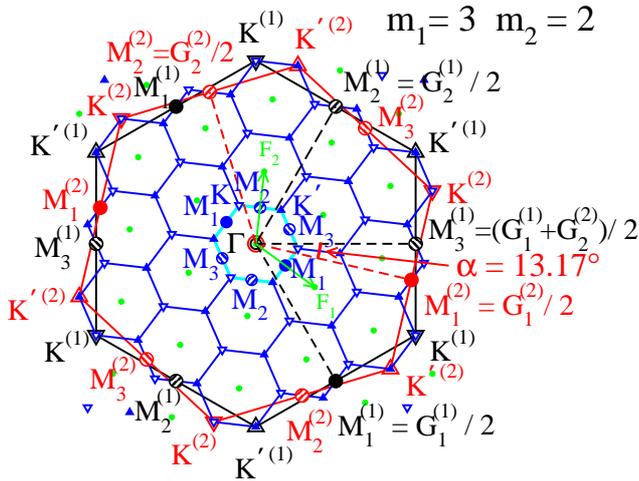}  
\caption{(Color online) 
Brillouin zone of the commensurately twisted bilayer graphene with $m_1=3$ and $m_2=2$. 
 The large black hexagon is the first Brillouin zone of the first layer, and the 
large red hexagon is the first Brillouin zone of the second layer.
Black  triangular-down's  (triangular-up's) are the K  (K$'$) points and the equivalent points of the
first layer, K$^{(1)}$ (K$^(1)'$). Red  triangular-down's  (triangular-up's) are  
the K  (K$'$) points and the equivalent points of the
first layer, K$^{(2)}$ (K$^(2)'$).  
The black circles (red circles) are the M$_1$, M$_2$, and M$_3$ points 
of the first layer (the second layer). 
The green small filled circles are the $\Gamma$ point and its equivalent points of the twisted bilayer graphene.
The cyan hexagon at the center is the first Brillouin zone 
of the twisted bilayer graphene.  
The open blue triangular-down's and filled triangular-up's are the K and K$'$ points 
(and their equivalent points) of the twisted bilayer graphene,
respectively.
M$_1$, M$_2$ and M$_3$ points are shown by the blue circles on the edge of the first Brillouin zone
of the twisted bilayer graphene.
}
\label{fig5twist43}
\end{figure}

Commensurately twisted bilayer graphene is labeled by two integers $m_1$ and $m_2$.
The primitive lattice vectors $\mathbf{L}_1$ and $\mathbf{L}_2$ of the commensurately twisted
bilayer graphene is given by\cite{Lopes2007,Moon2012,Hasegawa2013}
\begin{align}
 \mathbf{L}_1 &= m_1 \mathbf{a}_1^{(1)} + m_2 \mathbf{a}_2^{(1)} ,\\
 \mathbf{L}_2 &= R_{\frac{\pi}{3}} \mathbf{L}_1^{(1)} 
= -m_2 \mathbf{a}_1^{(1)} + (m_1+m_2) \mathbf{a}_2^{(1)} ,
\label{eqL2}
\end{align}
as shown in Fig.~\ref{figfig32}.
The number of sites in the unit cell of the commensurately rotated bilayer graphene is
\begin{equation}
4 n_0=4 (m_1^2+m_1m_2+m_2^2),
\end{equation}
in the supercell, i.e. $n_0$ A sites and B sites in each layer.

The $A$ sites are located at the positions
\begin{equation}
 m_1 \mathbf{a}_1^{(1)} + m_2 \mathbf{a}_2^{(1)} 
\end{equation}
with integer $m_1$ and $m_2$. 
We define the angle $\alpha$ by the angle between the vectors
\begin{equation}
 \mathbf{L}_1  = m_1 \mathbf{a}_1^{(1)} + m_2 \mathbf{a}_2^{(1)},  
 \end{equation}
 and
 \begin{equation}
 \mathbf{L}_1' = m_2 \mathbf{a}_1^{(1)} + m_1 \mathbf{a}_2^{(1)}.  
\end{equation}
Two vectors have the same length 
\begin{equation}
 \lvert \mathbf{L}_1 \rvert^2 =  \lvert \mathbf{L}_1 \rvert^2 = a^2 (m_1^2+m_1 m_2 + m_2^2 ) 
= a^2  n_0.
\end{equation}
By using  
\begin{equation}
 \mathbf{L}_1 \cdot \mathbf{L}_1' = \frac{a^2}{2} (m_1^2 + 4 m_1 m_2 + m_2^2),
\end{equation}
we obtain
\begin{equation}
  \cos \alpha=\frac{m_1^2+4 m_1 m_2 + m_2^2}{2 (m_1^2+m_1 m_2 + m_2^2)},
  \label{eqA13}
\end{equation}
and
\begin{equation}
  \sin \alpha=\frac{\sqrt{3}(m_1^2- m_2^2)}{2 (m_1^2+m_1 m_2 + m_2^2)},
  \label{eqA14}
\end{equation}
We can obtain the commensurate twisted bilayer graphene by rotating the second layer 
either $\alpha$ or $\pi/3 - \alpha$. In order to obtain the Bernal stacking (AB stacking) in
the limit $\alpha \to 0$, 
we study the $\pi/3 -\alpha$ rotation in the AA stacked bilayer graphene ($\alpha$ rotation
 in the Bernal stacked bilayer graphene) in this paper. 
Note that
\begin{align}
 \cos \left( \frac{\pi}{3} -\alpha \right) &= 
  \frac{2 m_1^2+2m_1 m_2 - m_2^2}{2(m_1^2+m_1 m_2 + m_2^2)} , \\
 \sin \left( \frac{\pi}{3} -\alpha \right) &= 
  \frac{\sqrt{3} m_2(2 m_1+m_2)}{2(m_1^2+m_1 m_2 + m_2^2)} .
\end{align}
The primitive lattice vectors of the second layer are  given by rotating $\pi/3 - \alpha$;
\begin{align}
 \mathbf{a}_1^{(2)} &= R_{\frac{\pi}{3}-\alpha} \mathbf{a}_1^{(1)} \nonumber \\
&= \frac{a}{2(m_1^2+m_1 m_2 + m_2^2)} \begin{pmatrix}
 \sqrt{3}m_1(m_1+ 2 m_2)\\ -m_1^2 +2 m_1 m_2 + 2 m_2^2 
\end{pmatrix},  \\
 \mathbf{a}_2^{(2)} &= R_{\frac{\pi}{3}-\alpha} \mathbf{a}_2^{(1)} \nonumber \\
 &= \frac{a}{2(m_1^2+m_1 m_2 + m_2^2)} \begin{pmatrix}
  \sqrt{3}(m_1^2-m_2^2) \\   m_1^2+4 m_1 m_2 + m_2^2
\end{pmatrix}.
\end{align}
%

%

The reciprocal Lattice vectors for the first layer ($\mathbf{G}_1^{(1)}$ and $\mathbf{G}_2^{(1)}$)
and the second layer ($\mathbf{G}_1^{(2)}$ and $\mathbf{G}_2^{(2)}$)
are the vectors in the wave-number space and satisfy the relations,
\begin{align}
 \mathbf{a}_i^{(\ell)} \cdot \mathbf{G}_j^{(\ell)} &= 2\pi \delta_{i,j}, \\
\intertext{and} 
 \hat{\mathbf{z}}  \cdot \mathbf{G}_i^{(\ell)}  &=0,
\end{align}
where $i, j, \ell=1$ or $2$ and $\hat{\mathbf{z}}$ is the unit vector along the $z$ direction.
They are given by
\begin{align}
  \mathbf{G}_1^{(\ell)} &= 2 \pi \frac{\mathbf{a}_2^{(\ell)} \times 
\hat{\mathbf{z}}}{ ( \mathbf{a}_1^{(\ell)} \times \mathbf{a}_2^{(\ell)} ) \cdot \hat{\mathbf{z}}},\\
  \mathbf{G}_2^{(\ell)} &= 2 \pi \frac{\hat{\mathbf{z}} \times \mathbf{a}_1^{(\ell)}  
}{ ( \mathbf{a}_1^{(\ell)} \times \mathbf{a}_2^{(\ell)} ) \cdot \hat{\mathbf{z}}}.
\end{align}
We obtain
\begin{align}
 \mathbf{G}_1^{(1)} &= \frac{2\pi}{a} \begin{pmatrix}
   \frac{1}{\sqrt{3}} \\ -1
 \end{pmatrix} \nonumber \\
  &= \frac{4\pi}{3a^2} (2 \mathbf{a}_1-\mathbf{a}_2 ) , \\
  \mathbf{G}_2^{(1)} &= \frac{2\pi}{a} \begin{pmatrix}
   \frac{1}{\sqrt{3}} \\ 1
 \end{pmatrix} \nonumber \\
 &= \frac{4\pi}{3a^2} (- \mathbf{a}_1 +2 \mathbf{a}_2 ) .
\end{align}
The reciprocal Lattice vectors for the second layer, $\mathbf{G}_1^{(2)}$ and $\mathbf{G}_2^{(2)}$
are obtained by rotating $\pi/3-\alpha$,
\begin{align}
 \mathbf{G}_1^{(2)} &= R_{\frac{\pi}{3}-\alpha} \mathbf{G}_1^{(1)}\nonumber \\
&=\frac{2\pi}{a} \frac{1}{m_1^2+m_1 m_2 + m_2^2} \begin{pmatrix}
   \frac{1}{\sqrt{3}} (m_1^2+4 m_1 m_2 +m_2^2) \\ -(m_1^2-m_2^2)
 \end{pmatrix},\\
  \mathbf{G}_2^{(2)} &=  R_{\frac{\pi}{3}-\alpha} \mathbf{G}_2^{(1)} \nonumber \\
&= \frac{2\pi}{a} \frac{1}{m_1^2+m_1 m_2 + m_2^2} \begin{pmatrix}
   \frac{1}{\sqrt{3}} (m_1^2-2 m_1 m_2 -2 m_2^2)\\ m_1 (m_1+2 m_2)
 \end{pmatrix}.
\end{align}

We write the reciprocal lattice vectors of the twisted bilayer graphene 
as $\mathbf{F}_1$ and $\mathbf{F}_2$, which are given by
\begin{align}
 &\mathbf{F}_1 = 2\pi 
\frac{\mathbf{L}_2 \times \hat{\mathbf{z}}}{(\mathbf{L}_1 \times \mathbf{L}_2) \cdot \hat{\mathbf{z}}},\\
 \mathbf{F}_2 &= 2\pi 
\frac{\hat{\mathbf{z}} \times \mathbf{L}_1}{(\mathbf{L}_1 \times \mathbf{L}_2) \cdot \hat{\mathbf{z}}}.
\end{align}
We obtain
\begin{align}
 &\mathbf{F}_1 = \frac{1}{m_1^2+m_1 m_2 + m_2^2} 
\left( (m_1+m_2) \mathbf{G}_1^{(1)} + m_2 \mathbf{G}_2^{(1)} \right) 
\label{eqF1}\\
&= \frac{4\pi}{3 a^2} \frac{ (2m_1+m_2) \mathbf{a}_1+
(-m_1+m_2) \mathbf{a}_2 }{m_1^2+m_1 m_2 + m_2^2} ,\\
 \mathbf{F}_2 &= \frac{1}{m_1^2+m_1 m_2 + m_2^2} 
\left( -m_2 \mathbf{G}_1^{(1)} + m_1 \mathbf{G}_2^{(1)} \right) .
 \label{eqF2} \\
 &= \frac{4\pi}{3 a^2} \frac{ (-m_1-2m_2) \mathbf{a}_1+
(2m_1+m_2) \mathbf{a}_2 }{m_1^2+m_1 m_2 + m_2^2} ,
\end{align}
We can write Eqs.~(\ref{eqF1}) and (\ref{eqF2}) as
\begin{align}
 \mathbf{G}_1^{(1)} &= m_1 \mathbf{F}_1 - m_2 \mathbf{F}_2, \\
 \mathbf{G}_2^{(1)} &= m_2 \mathbf{F}_1 + (m_1+m_2) \mathbf{F}_2.
\end{align}
We plot Brillouin zone in the twisted bilayer graphene with $m_1=3$ and $m_2=2$  in Fig.~\ref{fig5twist43}.

{In the continuous model, coupling between K$^{(1)}$ and three nearest K$'$ (including K$'^{(2)}$)
are taken into account.\cite{Bistritzer2011} 
On the other hand, the commensurately-twisted
bilayer graphene has an intervalley coupling between K$^{(1)}$ and K$^{(2)}$.
The intervalley coupling causes the band gap at K and K$'$, although 
it is small in the ambient pressure.\cite{Shallcross2008}
If $m_1-m_2=3 n$ with integer $n$, the band structure near K is drastically 
changed\cite{Sboychakov2015, Rozhkov2016}. In this paper we do not consider that case.
}

\bibliography{bib_69657}

\begin{thebibliography}{10}

\bibitem{Neto2009}
A.~H. Castro~Neto, F.~Guinea, N.~M.~R. Peres, K.~S. Novoselov, and A.~K. Geim:
  Rev. Mod. Phys. {\bfseries 81} (2009) 109.

\bibitem{Gibertini2009}
M.~Gibertini, A.~Singha, V.~Pellegrini, M.~Polini, G.~Vignale, A.~Pinczuk,
  L.~N. Pfeiffer, and K.~W. West: Phys. Rev. B {\bfseries 79} (2009) 241406.

\bibitem{Downing2017}
C.~A. Downing and M.~E. Portnoi: Journal of Physics: Condensed Matter
  {\bfseries 29} (2017) 315301.

\bibitem{Rozhkov2016}
A.~V. Rozhkov, A.~O. Sboychakov, A.~L. Rakhmanov, and F.~Nori: Phys. Rep.
  {\bfseries 648} (2016) 1.

\bibitem{Lopes2007}
J.~M.~B. Lopes~dos Santos, N.~M.~R. Peres, and A.~H. Castro~Neto: Phys. Rev.
  Lett. {\bfseries 99} (2007) 256802.

\bibitem{Bistritzer2011}
R.~Bistritzer and A.~H. MacDonald: Proceedings of the National Academy of
  Sciences {\bfseries 108} (2011) 12233.

\bibitem{Morell2010}
E.~Su\'arez~Morell, J.~D. Correa, P.~Vargas, M.~Pacheco, and Z.~Barticevic:
  Phys. Rev. B {\bfseries 82} (2010) 121407.

\bibitem{Po2018}
H.~C. Po, L.~Zou, A.~Vishwanath, and T.~Senthil: Phys. Rev. X {\bfseries 8}
  (2018) 031089.

\bibitem{Hejazi2019}
K.~Hejazi, C.~Liu, H.~Shapourian, X.~Chen, and L.~Balents: Phys. Rev. B
  {\bfseries 99} (2019) 035111.

\bibitem{Tarnopolsky2019}
G.~Tarnopolsky, A.~J. Kruchkov, and A.~Vishwanath: Phys. Rev. Lett. {\bfseries
  122} (2019) 106405.

\bibitem{Cao2018}
Y.~Cao, V.~Fatemi, A.~Demir, S.~Fang, S.~L. Tomarken, J.~Y. Luo, J.~D.
  Sanchez-Yamagishi, K.~Watanabe, T.~Taniguchi, E.~Kaxiras, R.~C. Ashoori, and
  P.~Jarillo-Herrero: Nature {\bfseries 556} (2018) 80.

\bibitem{Cao2018b}
Y.~Cao, V.~Fatemi, S.~Fang, K.~Watanabe, T.~Taniguchi, E.~Kaxiras, and
  P.~Jarillo-Herrero: Nature {\bfseries 556} (2018) 43.

\bibitem{Carr2018}
S.~Carr, S.~Fang, P.~Jarillo-Herrero, and E.~Kaxiras: Phys. Rev. B {\bfseries
  98} (2018) 085144.

\bibitem{Yankowitz2019}
M.~Yankowitz, S.~Chen, H.~Polshyn, Y.~Zhang, K.~Watanabe, T.~Taniguchi,
  D.~Graf, A.~F. Young, and C.~R. Dean: Science {\bfseries 363} (2019) 1059.

\bibitem{Shallcross2008}
S.~Shallcross, S.~Sharma, and O.~A. Pankratov: Phys. Rev. Lett. {\bfseries 101}
  (2008) 056803.

\bibitem{Mele2010}
E.~J. Mele: Phys. Rev. B {\bfseries 81} (2010) 161405.

\bibitem{Shallcross2010}
S.~Shallcross, S.~Sharma, E.~Kandelaki, and O.~A. Pankratov: Phys. Rev. B
  {\bfseries 81} (2010) 165105.

\bibitem{Uchida2014}
K.~Uchida, S.~Furuya, J.-I. Iwata, and A.~Oshiyama: Phys. Rev. B {\bfseries 90}
  (2014) 155451.

\bibitem{Sboychakov2015}
A.~O. Sboychakov, A.~L. Rakhmanov, A.~V. Rozhkov, and F.~Nori: Phys. Rev. B
  {\bfseries 92} (2015) 075402.

\bibitem{Rozhkov2017}
A.~V. Rozhkov, A.~O. Sboychakov, A.~L. Rakhmanov, and F.~Nori: Phys. Rev. B
  {\bfseries 95} (2017) 045119.

\bibitem{Wolf2019}
T.~M.~R. Wolf, J.~L. Lado, G.~Blatter, and O.~Zilberberg: Phys. Rev. Lett.
  {\bfseries 123} (2019) 096802.

\bibitem{Hasegawa2006}
Y.~Hasegawa and M.~Kohmoto: Phys. Rev. B {\bfseries 74} (2006) 155415.

\bibitem{Esaki2009}
K.~Esaki, M.~Sato, M.~Kohmoto, and B.~I. Halperin: Phys. Rev. B {\bfseries 80}
  (2009) 125405.

\bibitem{Hasegawa2012}
Y.~Hasegawa and K.~Kishigi: Phys. Rev. B {\bfseries 86} (2012) 165430.

\bibitem{Popov2011}
A.~M. Popov, I.~V. Lebedeva, A.~A. Knizhnik, Y.~E. Lozovik, and B.~V. Potapkin:
  Phys. Rev. B {\bfseries 84} (2011) 045404.

\bibitem{Nam2017}
N.~N.~T. Nam and M.~Koshino: Phys. Rev. B {\bfseries 96} (2017) 075311.

\bibitem{Moon2012}
P.~Moon and M.~Koshino: Phys. Rev. B {\bfseries 85} (2012) 195458.

\bibitem{Hasegawa2013}
Y.~Hasegawa and M.~Kohmoto: Phys. Rev. B {\bfseries 88} (2013) 125426.

\end{thebibliography}

\end{document}